\documentclass[a4paper,11pt]{article}
\usepackage{amsmath}
\usepackage{graphicx}
\usepackage{amssymb}
\usepackage{cite}
\usepackage{natbib}
\usepackage[hidelinks]{hyperref}

\hypersetup{
  colorlinks = true,     
  urlcolor = blue,       
  linkcolor = blue,      
  citecolor = red         
}

\renewcommand{\vec}[1]{\mathbf{#1}}

\newcommand{\rmi}{\mathrm{i}}
\newcommand{\rme}{\mathrm{e}}

\newcommand{\bra}[1]{\left<#1\right|}
\newcommand{\ket}[1]{\left|#1\right>}

\newcommand{\iprod}[2]{\left<#1|#2\right>}

\providecommand{\keywords}[1]
{
  \small	
  \textbf{\textit{Keywords---}} #1
}

\begin{document}

\title{Time evolution of quantum gates and the necessity of complex numbers}

\author{M.P. Vaughan\footnote{\emph{Corresponding author}: martin.vaughan@physics.org}}

\date{\small{Version: 1.1}}

\maketitle

\begin{abstract}
As physical systems, qubits cannot be transformed instantaneously but must evolve from some input state to their output state. We describe a simple scheme in which the effect of a quantum gate is described by the action of an effective Hamiltonian acting for some characteristic time $\tau$. We use this method to model the temporal dynamics of some common quantum logic gates and highlight the role of the complex phase in the time evolution of these systems. This model shows that the action of common unary gates is to induce Bloch sphere trajectories along lines of latitude relative to an eigenvector of the gate. Such trajectories would immediately move a `rebit', initially confined to a line of latitude, off this line and acquire a complex phase. The role of the complex phase in bringing about the entanglement of two qubits is is also highlighted. It is then asked whether such dynamics could be modelled using real quantum mechanics. It is shown that the continuous evolution required for such dynamics can only be provided by members of the special orthogonal group of the vector space. Since the matrices representing many quantum gates of interest have determinant -1, no real special orthogonal operators can model their evolution if the dimension of the real vector space is the same as that of the complex space. Next we look at the mapping from a complex vector space of dimension $N$ to a real space of dimension $2N$ that is often used to construct `real' quantum mechanics. It is shown that this is just an isomorphic mapping from the scalar representation of complex numbers to their $2\times2$ matrix equivalents, so that the resulting matrices are actually representations of complex matrices. Finally, we investigate the endomorphism of real vector spaces of dimension $N = 2^{n}$, where $n \in \mathbb{Z}^{+}$, suitable for the modelling of $n-1$ qubits. We confirm that the mapping from $\mathbb{C}^{2^{n-1}} \to \mathbb{R}^{2^{n}}$ only maps elements of $\mathrm{End}(\mathbb{C}^{2^{n-1}})$ to a restricted subspace of $\mathrm{End}(\mathbb{R}^{2^{n}})$ that reproduces the `real' representation of complex matrices.     
\end{abstract}



\section{Introduction}
In addition to the profound technological applications it proffers, quantum information theory provides a fertile ground for the investigation of fundamental questions in quantum theory. One such perennial question is whether complex numbers are actually needed to frame quantum theory properly. A recent Nature article by Renou \emph{et al}~\cite{renou2021quantum} has argued that the predictions of real-valued formulations may be discriminated from those of the conventional complex theory experimentally. What actually constitutes a `real' formulation of quantum mechanics, however, may in itself be a contentious question. 

In this article we shall consider two variants on this theme. Firstly, we consider the idea of `rebits'\cite{caves2001entanglement, rudolph2002} - real valued versions of qubits, which represent the binary quantum units of information. These have primarily been introduced as a foil theory to compare and contrast the predictions of real vs complex formulations of quantum mechanics. In this paper, we focus specifically on the time-dependence of quantum systems and investigate what the consequences of this might be when we try to constrain the qubits to take only real values.

More generally, a common scheme used in the literature~\cite{myrheim1999quantum, aleksandrova2013real} is a mapping from a complex Hilbert space to a real one. For finite dimensional spaces this always involves a doubling of the dimensions used. However, as we shall argue in this paper, such mappings often just represent the original complex space using real numbers due to the fact that they are mapping scalar complex elements to their $2\times2$ matrix equivalents. We would therefore claim that calling this a `real' formulation of quantum mechanics is incorrect and misleading. Nonetheless, we shall investigate both of these topics in the specific context of quantum logic gate dynamics.

\subsection{Rebits vs qubits}
In order to define what we mean by a rebit, let us first recall how we specify a qubit. A qubit is a state of a 2D complex Hilbert space representing a binary unit of information. Expressing a qubit in the general form

\begin{align}
\ket{\theta, \phi} = \left[\begin{array}{c}
\cos(\theta/2) \\
\sin(\theta/2)\rme^{\rmi\phi}\end{array}\right], \label{eq:general-qubit}
\end{align}

\noindent where the parameters $\theta$ and $\phi$ correspond to polar coordinates, the qubit can be mapped to a point on the Bloch sphere. In general, the components of a qubit will be complex but one could imagine constraining them to be real. Such a rebit would then lie on a line of longitude on the Bloch sphere - for example the great circle defined by $\phi=0$, which passes through the points representing $\ket{0} = [1~0]^{T}$ (at $\theta=0$) and that representing $\ket{X_{0}} = [1~1]^{T}/\sqrt{2}$ (at $\theta=\pi/2)$.

If, in addition, we constrain ourselves to operations that involve only `real' quantum logic gates (i.e. gates such as the NOT or Hadamard gate that can be specified using real numbers), we might imagine that any computational process will only take us from one point on the line of longitude to another. If this were the case, then we have envisaged some form of quantum computing without complex numbers. A principal result of this paper is to show that this is not possible.  

\subsection{Overview}
In quantum logic circuits, qubits are transformed by the actions of logic gates, modelled in the complex theory by unitary operators. However, in reality, qubits and the gates that act on them are \emph{physical systems}, so such transformations do not happen instantaneously. As physical systems, quantum gates must cause qubits to evolve from their input states to their output states and these dynamics should be described by quantum theory. 

In this paper, we investigate this evolution by considering that a unitary operator representing a gate can be modelled by an effective Hamiltonian acting on the qubit for some characteristic time. We give the general mathematics for this in Section~\ref{eq:general-time-dependence} before solving the explicit time-dependence of some common unary logic gates in Section~\ref{sec:ops-on-C2}, using the standard complex theory. We find in this section that the evolution of these gates can be described by Bloch sphere trajectories along \emph{lines of latitude} relative to an axis defined by the eigenvectors of the unitary operator modelling the gate. Hence, a rebit confined to a line of longitude will be immediately moved off this line and will acquire an imaginary component. So, even if a Bloch sphere trajectory returns to a point on the line of longitude the state started from (where its components are real), the intermediate states will be complex.

In Section~\ref{sec:entanglement}, we move to systems of 2 qubits and discuss the emergence of entanglement. This is again a physical, evolutionary process that depends on the components of the system interacting with one another. Here, we continue to use the standard complex formulation in order to highlight the role of the complex phase in bringing about entanglement. Using the procedure of Section~\ref{eq:general-time-dependence}, we specifically model the time-evolution of a CNOT gate and give the preparation of Bell states as an example. 

It may be argued that the conclusion of Sections~\ref{sec:ops-on-C2} and \ref{sec:entanglement}, that time evolution requires complex numbers, suffers from circularity, since the results were derived using the standard complex theory. So, in Section~\ref{sec:ops-on-RN}, we consider how we might model the continuous evolution of a state using orthogonal operators on a real $N$-dimensional Hilbert space $\mathbb{R}^{N}$. It is shown there that, for \emph{continuous} evolution, we actually require the \emph{special orthogonal} operators of the $SO(N)$ group. We then show that these cannot be used to model the common unary logic gates in $\mathbb{R}^{2}$ nor the binary gates of $\mathbb{R}^{4}$ as the matrices representing them are not members of this group.

Whilst modelling the continuous evolution of logic gates is not possible in $\mathbb{R}^{2}$, it is common practice when formulating quantum mechanics on a real Hilbert space to map the elements of complex $N$-dimensional space $\mathbb{C}^{N}$ to a real vector space with twice the dimensions $\mathbb{R}^{2N}$. We show in Section~\ref{sec:mapping-CN-R2N} that this mapping actually just connects $\mathbb{C}^{N}$ to a real space representation of a complex vector space via the isomorphism between the scalar form of a complex number and its $2\times2$ matrix equivalent. In other words, the mapping only \emph{looks} like a real space formulation - each $2\times2$ element of a matrix in $\mathbb{R}^{2N}$ still represents a complex number.

Finally, in Section~\ref{sec:antisym-R4}, we address the question of whether real space operators that can model continuous time evolution are \emph{always} a representation of complex operators. We do this by investigating the endomorphism of real vector spaces of dimension $N = 2^{n}$, which would be suitable for modelling systems of $n-1$ qubits. We see there that the antisymmetric basis elements of $\mathrm{End}(\mathbb{R}^{2^{n}})$ (required for constructing members of the special orthogonal group) may always be considered to be representations of complex matrices. A general antisymmetric matrix, though, may not be consistently interpreted in this way. Moreover, many of the symmetric operators of $\mathrm{End}(\mathbb{R}^{2^{n}})$ cannot be interpreted as being complex at all.

 However, it is shown that the mapping from a complex to a real vector space described in Section~\ref{sec:mapping-CN-R2N} does not produce a one-to-one relationship between the basis elements of the two endomorphisms and many of the operators of $\mathrm{End}(2^{n})$ remain unused. It is then reaffirmed that this mapping \emph{does} consistently map a complex vector space to its real representation and so still constitutes a \emph{complex} vector space.  

\section{General time dependence of a unitary operator on $\mathbb{C}^{N}$}\label{eq:general-time-dependence}
A unitary operator $\vec{U}$ on an $N$-dimensional complex Hilbert space $\mathbb{C}^{N}$ may be taken to be due to the action of an effective Hamiltonian $\vec{H}$ acting for some characteristic time $\tau$ via

\begin{align}
\vec{U} &= \rme^{-\rmi \vec{H}\tau/\hbar}. \nonumber
\end{align}

\noindent It can then be shown that the eigenvectors of the $\vec{U}$ will be exactly those of $\vec{H}$. The eigenvalues of a unitary matrix, as is well-known, are all of modulus unity and so will be of the form

\begin{align}
\lambda_{i} = \rme^{\rmi \varphi_{i}}, \nonumber
\end{align}

\noindent for some real number $\varphi_{i}$. These will be related to the energy eigenvalues $E_{i} = \hbar\omega_{i}$ of the Hamiltonian $\vec{H}$ by

\begin{equation}
\omega_{i} = - \frac{\varphi_{i}}{\tau}. \nonumber
\end{equation}

We may then express the evolution of the unitary matrix due to $\vec{H}$ at some arbitrary time $t$ as

\begin{align}
\vec{U}(t) &= \sum_{i} \rme^{\rmi \varphi_{i}t/\tau} \ket{\phi_{i}}\bra{\phi_{i}}, \label{eq:general-time-dependence}
\end{align}

\noindent As is readily seen, at $t= 0$ this reproduces the identity matrix $\vec{I}$ on $\mathbb{C}^{N}$

\begin{align}
\vec{U}(0) &= \sum_{i} \ket{\phi_{i}}\bra{\phi_{i}} = \vec{I}. \nonumber
\end{align}

\noindent At time $t=\tau$, \eqref{eq:general-time-dependence} will have evolved to

\begin{align}
\vec{U}(\tau) &= \sum_{i} \rme^{\rmi \varphi_{i}} \ket{\phi_{i}}\bra{\phi_{i}} = \vec{U}, \nonumber
\end{align}

\noindent which is just the spectral decomposition of $\vec{U}$.

Now, in canonical form, the spectral decomposition of a unitary matrix $\vec{U}_{A}$ can be written down as

\begin{align}
\vec{U}_{A}\boldsymbol{\Phi}_{A} &= \boldsymbol{\Phi}_{A}\boldsymbol\Lambda, \nonumber
\end{align}

\noindent where $\boldsymbol\Lambda$ is a diagonal matrix of the eigenvalues of $\vec{U}_{A}$ and the columns of $\boldsymbol{\Phi}_{A}$ are the corresponding normalised eigenvectors. If a second matrix $\vec{U}_{B}$ shares exactly the same eigenvalues as $\vec{U}_{A}$, then we can find the time dependence of the second operator directly from

\begin{align}
\vec{U}_{B}(t) &= \boldsymbol{\Phi}_{B}\boldsymbol{\Phi}_{A}^{\dagger}\vec{U}_{A}(t)\boldsymbol{\Phi}_{A}\boldsymbol{\Phi}_{B}^{\dagger}. \label{eq:common-eigs}
\end{align}

\noindent This result will prove to be useful in the analysis of the time-dependence of quantum logic gates, which often have the same eigenvalues.

Obviously the procedure described here is just a model simplification. In reality, the actual processes by which a qubit is transformed may be much more complicated than can be described by \eqref{eq:general-time-dependence} alone. Indeed, if we combined several gates together, we could calculate a \emph{single} combined gate that could be solved by assuming just one Hamiltonian $\vec{H}$ acting for a particular time $\tau$. This would give the same end result as each individual gate $\vec{U}_{i}$ acting under its own Hamiltonian $\vec{H}_{i}$ for its own characteristic time $\tau_{i}$. The point here is that any physical system must evolve in \emph{some} way and that, for simple enough systems, we can model this using \eqref{eq:general-time-dependence}.    

\section{Complex unitary operators on $\mathbb{C}^{2}$}\label{sec:ops-on-C2}

In this section, we consider the time evolution of some common unary quantum logic gates, which can all be analysed on the 2D Hilbert space $\mathbb{C}^{2}$. As possibly the easiest to analyse, we shall begin with the $\vec{Z}$ gate (i.e the Pauli matrix $\sigma_{z}$). As a diagonal matrix, its eigenvalues $\pm 1$ are simply read from its diagonal elements. Moreover, since the eigenvectors of $\vec{Z}$ make up the identity matrix on $\mathbb{C}^{2}$, the result of \eqref{eq:common-eigs} reduces to

\begin{align}
\vec{U}_{\pm1}(t) &= \boldsymbol{\Phi}_{\pm1}\vec{Z}(t)\boldsymbol{\Phi}_{\pm1}^{\dagger}, \label{eq:from-Z}
\end{align}

\noindent where $\vec{U}_{\pm1}$ is any other $2\times2$ gate with eigenvalues of $\pm1$.

\subsection{The $\vec{Z}$, $\vec{T}$ and $\vec{S}$ gates}
Since the corresponding $\varphi_{i}$ values for eigenvalues of $\pm1$ are $0$ and $\pi$, we can write down the time evolution of $\vec{Z}$ as

\begin{align}
\vec{Z}(t) = \left[\begin{array}{cc}
1 & 0 \\
0 & \rme^{\rmi\pi t/\tau}\end{array}\right]. \label{eq:Z-time}
\end{align}

\noindent Acting on the  general qubit of \eqref{eq:general-qubit}, we then have

\begin{align}
\vec{Z}(t)\ket{\theta, \phi} &= \left[\begin{array}{c}
\cos(\theta/2) \\
\sin(\theta/2)\rme^{\rmi(\phi + \pi t/\tau)}\end{array}\right] = \ket{\theta, \phi+\pi t/\tau}. \label{eq:Zt-action}
\end{align}

\noindent This describes a state traversing around the Bloch sphere \emph{along a line of latitude} in a plane normal to the polar axis. So, even if $\ket{\theta, \phi}$ starts as a rebit on a line of longitude, it will immediately move off of this line and acquire an imaginary component under the action of $\vec{Z}(t)$. Note that at $t = \tau$, the state will have moved to the opposite side of the Bloch sphere and onto the line of longitude it started from. For example, an $\ket{X_{0}}$ state (an eigenvector of the Pauli $\sigma_{x}$ matrix) will have evolved to  $\ket{X_{1}}$ (its orthogonal complement). After a period of $2\tau$, the state will have undergone a complete traversal of the Bloch sphere and returned to its original state. This is reflected by the fact that $\vec{Z}^{2} = \vec{I}$.

At intermediate times, $\vec{Z}(t)$ reproduces the actions of two other quantum gates.  $\vec{Z}(\tau/4)$ gives us the \textbf{T} gate, whilst $\vec{Z}(\tau/2)$ produces the \textbf{S} or \emph{phase} gate. Here we have an example where a series of gates \emph{can} be modelled by combining repeated use of \eqref{eq:general-time-dependence}. This is possible in this case because all the gates in question have the same eigenvectors and can be modelled by the same effective Hamiltonian acting for different characteristic times.

\subsection{The $\vec{X}$ gate}
In common with the $\vec{Z}$ gate, the $\vec{X}$, $\vec{Y}$ and Hadamard gates all  have eigenvalues of $\pm1$, so we may use the prescription of \eqref{eq:from-Z} to derive their time dependence. For the $\vec{X}$ or NOT gate (equivalent to the Pauli $\sigma_{x}$ matrix), we can write the eigenvector matrix as

\begin{align}
\boldsymbol{\Phi}_{X} = \left[\begin{array}{cc}
\cos(\pi/4) & -\sin(\pi/4) \\
\sin(\pi/4) & \cos(\pi/4)\end{array}\right], \label{eq:phi-X}
\end{align}

\noindent where we have multiplied the usual expression for $\ket{X_{1}}$ by a global phase factor of $-1$ for mathematical convenience. In this form, it is easy to see that $\boldsymbol{\Phi}_{X}$ corresponds to a rotation of $\pi/4$ radians. On the Bloch sphere, this corresponds to a rotation of $\theta = \pi/2$, so essentially rotates the $z$-axis to point along the $x$ direction. Using \eqref{eq:from-Z}, we get

\begin{align}
\vec{X}(t) &= \frac{1}{2}\left[\begin{array}{cc}
1 + \rme^{\rmi\pi t/\tau} & 1 - \rme^{\rmi\pi t/\tau} \\
1 - \rme^{\rmi\pi t/\tau} & 1 + \rme^{\rmi\pi t/\tau}\end{array}\right]. \label{eq:X-time}
\end{align}

\noindent The Bloch sphere trajectories in this case are similar to those for the $\vec{Z}(t)$ gate, except this time they orbit around the axis defined by the  $\ket{X_{0}}$ vector. We will see that this is a common feature: the Bloch sphere trajectories of all $\pm1$ eigenvalue gates are orbits along a line of latitude relative to an axis defined by the `0' eigenvector of the gate.

\subsection{The $\vec{Y}$ gate}
The eigenvector matrix for the $\vec{Y}$ gate (Pauli $\sigma_{y}$ matrix) may be written as

\begin{align}
\boldsymbol{\Phi}_{X} &=  \left[\begin{array}{cc}
1 & 0 \\
0 & \rme^{\rmi\pi/2}\end{array}\right]\left[\begin{array}{cc}
\cos(\pi/4) & -\sin(\pi/4) \\
\sin(\pi/4) & \cos(\pi/4)\end{array}\right], \label{eq:phi-Y}
\end{align}

\noindent where, again, we have multiplied the usual expression for $\ket{Y_{1}}$ by $-1$. This may then be seen as the same Bloch sphere rotation as \eqref{eq:phi-X}, for the $\vec{X}$ gate, following by a rotation about the $z$-axis by $\pi/2$ radian. In other words, a transformation of the $z$-axis to point along the $y$-direction. We then find the temporal evolution

\begin{align}
\vec{Y}(t) = \frac{1}{2}\left[\begin{array}{cc}
1 + \rme^{\rmi\pi t/\tau} & -\rmi(1 - \rme^{\rmi\pi t/\tau}) \\
\rmi(1 - \rme^{\rmi\pi t/\tau}) & 1 + \rme^{\rmi\pi t/\tau}\end{array}\right]. \label{eq:Y-time}
\end{align}

\subsection{The Hadamard gate}

The Hadamard or `$H$' gate (here, denoted by $\vec{U}_{H}$ to avoid confusion with a Hamiltonian $\vec{H}$) is of particular interest in quantum information theory due to its use in 2-qubit circuits to generate Bell states. Its matrix is given by

\begin{align}
\vec{U}_{H} &= \frac{1}{\sqrt{2}}\left[\begin{array}{cc}
1 & 1 \\
1 & -1\end{array}\right], \label{eq:hadamard-gate}
\end{align}

\noindent which has an eigenvector matrix

\begin{align}
\ket{\phi_{+}} &= \left[\begin{array}{cc}
\cos(\pi/8)  & -\sin(\pi/8) \\
\sin(\pi/8) & \cos(\pi/8) \end{array}\right]. \label{eq:phi-H}
\end{align}

\noindent The first column of this, $\ket{H_{0}}$, then corresponds to a point on the great circle passing through both $\ket{0}$ and $\ket{X_{0}}$ with a polar coordinate of $\theta = \pi/4$. Again we see that eigenvector matrix represents a transformation to align the $z$-axis with this vector. 

Putting $c = \cos(\pi/8)$ and $s = \sin(\pi/8)$ for brevity, the time evolution for the gate is

\begin{align}
\vec{U}_{H}(t) &= \left[\begin{array}{cc}
c^{2} + s^{2}\rme^{\rmi\pi t/\tau}  & sc(1 - \rme^{\rmi\pi t/\tau}) \\
sc(1 - \rme^{\rmi\pi t/\tau}) & s^{2} + c^{2}\rme^{\rmi\pi t/\tau}\end{array}\right]. \label{eq:hadamard-time}
\end{align}

\noindent Once more, the Bloch sphere trajectories are orbits along lines of latitude relative to the vector $\ket{H_{0}}$.

\subsection{Consequences for rebits on $\mathbb{C}^{2}$}
So far, we have used conventional, complex quantum mechanics to derive the time-evolution of some common unary quantum logic gates. We have seen that, even for apparently real gates such as the $\vec{Z}$, $\vec{X}$ and $\vec{U}_{H}$ gates, the time evolution requires complex numbers for its description. Such evolution is characterised by trajectories following lines of latitude relative to an eigenvector of the gate. This means that, under the representation of such gates in $\mathbb{C}^{2}$, we cannot have rebits constrained to lie on a line of longitude of the Bloch sphere, even if the action of the gates is to map them from one point on that line of longitude to another.

This is perhaps not too surprising, since we have, after all, been using complex quantum mechanics to describe the time evolution. The question then remains whether it is possible to describe such time evolution in a \emph{real} Hilbert space. In order to address this question, we must first consider some general properties of orthogonal operators in such spaces, which we shall investigate in the Section~\ref{sec:ops-on-RN}. Before doing so, we should first look at the emergence of entanglement from the point of view of complex quantum mechanics.

\section{Entanglement}\label{sec:entanglement}
\subsection{Entanglement requires interaction}
Moving on to systems of two qubits, we encounter the quintessentially quantum mechanical feature of entanglement. In the literature, comparisons between rebits and qubits tend to focus on the differences between the measures of entropy, particularly for mixed states, such as the entanglement of formation~\cite{caves2001entanglement}. What tends not to be treated is the question of how systems become entangled in the first place. It is straight-forward to show that a bipartite system consisting of subsystems $A$ and $B$ initially in a tensor product state will remain unentangled if acted on only by the self-Hamiltonians of these systems. 

In order to become entangled (or for an entangled system to become unentangled) the subsystems must \emph{interact} with each other. This needs to be described by an \emph{interaction Hamiltonian} $\vec{H}_{I}$ that acts on \emph{both} subsystems. This in turn implies the system must \emph{evolve} from a tensor product to an entangled state. In this section, we continue to analyse this in terms of conventional complex quantum mechanics in order to highlight the role played by the complex phase in bringing this about.

Let us assume a bipartite system initially in a tensor product

\begin{align}
\ket{\Psi_{0}} &= \ket{\psi_{0}^{A}}\otimes\ket{\psi_{0}^{B}} \equiv \ket{\psi_{0}^{A}\psi_{0}^{B}}. \label{eq:tensor-product}
\end{align} 

\noindent For simplicity, we shall assume that the only Hamiltonian acting on the system will be the interaction Hamiltonian $\vec{H}_{I}$, which may be expressed in the form

\begin{align}
\vec{H}_{I} &= \sum_{ij} \hbar\omega_{ij}\ket{\phi_{ij}}\bra{\phi_{ij}}. \label{eq:interaction-H}
\end{align}

\noindent Here, the double indices imply summation over the degrees of freedom of both subsystems and $\ket{\phi_{ij}}$ is a joint state on the total system. In many cases (although unfortunately not for some interesting 2-qubit gates) the state $\ket{\phi_{ij}}$ can be factorised into states of the $A$ and $B$ subsystems. The crucial aspect here that brings about the entanglement is that the energy eigenvalues $\hbar\omega_{ij}$ are for the \emph{joint} system. 

With only $\vec{H}_{I}$ acting, a unitary operator describing the time evolution of the joint system can then be formulated as

\begin{align}
\vec{U}(t) &= \rme^{-\rmi\vec{H}_{I}t/\hbar}, \nonumber \\
&= \sum_{ij} \rme^{-\rmi\omega_{ij}t} \ket{\phi_{ij}}\bra{\phi_{ij}}. \label{eq:2-qubit-time-dependence}
\end{align}

\noindent The time-dependent solution for the bipartite system is simply solved as

\noindent 
\begin{align}
\ket{\Psi(t)} &= \sum_{ij} \rme^{-\rmi\omega_{ij}t} \ket{\phi_{ij}}\iprod{\phi_{ij}}{\Psi_{0}}. \label{eq:evolve-entangle}
\end{align} 

\noindent Note that, by assumption, at $t = 0$, this system is in a tensor product. As it evolves, however, it \emph{may} become entangled. From the form of \eqref{eq:evolve-entangle} it is clear to see how this can happen. The phase factors $\rme^{-\rmi\omega_{ij}t}$ coupling the two subsystems cannot, in general, be factorised (except when they take the form $\omega_{ij} = \omega_{i} + \omega_{j}$). What we see here is that, under unitary time evolution, entanglement emerges \emph{solely} through the effect of the complex phase coupling the two subsystems. 

\subsection{The CNOT gate}\label{sec:CNOT}
As a pertinent example of a bipartite system, let us consider the CNOT gate, which we shall denote by $\vec{U}_{\bar{C}}$. Using the general prescription of Section~\ref{eq:general-time-dependence}, we can derive its time evolution by assuming a Hamiltonian $\vec{H}_{I}$ acting for time $\tau$.

The eigenstates of the  $\vec{U}_{\bar{C}}$ are

\begin{align}
\ket{\phi_{00}} &= \ket{00},~~\ket{\phi_{01}} = \ket{01},  \nonumber \\
\ket{\phi_{10}} &= \frac{1}{\sqrt{2}}\left(\ket{10} + \ket{11}\right),~~\ket{\phi_{11}} = \frac{1}{\sqrt{2}}\left(\ket{11} - \ket{10}\right). \nonumber
\end{align}

\noindent Relating the eigenvalues of the gate to the corresponding energy eigenvalues, we find $\omega_{00} = \omega_{01} = \omega_{10} = 0$ and $\omega_{11} = -\pi/\tau$. The time-dependence is then found to be

\begin{align}
\vec{U}_{\bar{C}}(t) = \left[\begin{array}{cccc}
1 & 0 & 0 & 0 \\
0 & 1 & 0 & 0 \\
0 & 0 & \tfrac{1}{2}(1 + \rme^{\rmi\pi t/\tau}) & \tfrac{1}{2}(1 - \rme^{\rmi\pi t/\tau}) \\
0 & 0 & \tfrac{1}{2}(1 - \rme^{\rmi\pi t/\tau}) & \tfrac{1}{2}(1 + \rme^{\rmi\pi t/\tau}) \end{array}\right]. \label{eq:CNOT}
\end{align}

\noindent The reader can easily check that this starts off as the $4\times4$ identity matrix at $t=0$ and evolves to the CNOT gate at $t=\tau$.

For an interesting application of the CNOT gate, we can prepare an input state $\ket{00}$ by acting on the first qubit with the Hadamard gate to get

\begin{align}
\ket{\Psi_{0}} &= (\vec{U}_{H}\otimes\vec{I}_{2})\ket{00} = \frac{1}{\sqrt{2}}\left(\ket{00} + \ket{10}\right). 
\end{align} 

\noindent Evolving this with $\vec{U}_{\bar{C}}(t)$, we then have

\begin{align}
\ket{\Psi(t)} &= \frac{1}{\sqrt{2}}\left(\ket{00} + \tfrac{1}{2}(1 + \rme^{\rmi\pi t/\tau})\ket{10} + \tfrac{1}{2}(1 - \rme^{\rmi\pi t/\tau})\ket{11}\right). 
\end{align} 

\noindent At $t=\tau$, $\ket{\Psi(t)}$ has evolved to one of the Bell states

\begin{align}
\ket{\Psi(\tau)} &= \frac{1}{\sqrt{2}}\left(\ket{00} + \ket{11}\right). 
\end{align} 

\noindent We can obtain the other Bell states by starting with the other elements of the standard basis for a two qubit system $\ket{01}$, $\ket{10}$ and $\ket{11}$.

\section{Real orthogonal operators on $\mathbb{R}^{N}$}\label{sec:ops-on-RN}

\subsection{Evolution operators on $\mathbb{R}^{N}$ must belong to $SO(N)$}\label{sec:evolve-RN}
The analogue of a unitary operator on a real Hilbert space is an \emph{orthogonal} operator having the property

\begin{align}
\tilde{\vec{U}}^{T}\tilde{\vec{U}} &= \tilde{\vec{U}}\tilde{\vec{U}}^{T} = \vec{I}. \nonumber
\end{align}

\noindent (Here we adopt the convention of indicating a general matrix constructed on a real Hilbert space with an overlying tilde symbol `$\sim$'. For specific matrices common to both real and complex spaces, we will often drop this). Similarly to unitary operators, orthogonal operators also preserve the length of a vector they operate on and so are crucial for preserving the normalisation of that vector. In a real Hilbert space, then, it is orthogonal operators that we must look to in order to provide the mechanics of temporal evolution.

Since we model time, $t$, as a continuous parameter, it is essential that a time evolution operator $\tilde{\vec{U}}(\delta t)$ can be defined for an arbitrarily small change in $t$. For such an infinitesimal time increment $\delta t$ we should then have 

\begin{align}
\tilde{\vec{U}}(\delta t) &\approx \vec{I} + \tilde{\boldsymbol\Omega}\delta t, \nonumber
\end{align}

\noindent for some real matrix $\tilde{\boldsymbol\Omega}$ to be determined. The orthogonality condition on $\tilde{\vec{U}}(t)$ then implies that, for vanishing $\delta t$, we must have

\begin{align}
\tilde{\boldsymbol\Omega}^{T} &= - \tilde{\boldsymbol\Omega}. \nonumber
\end{align}

\noindent That is, $\tilde{\boldsymbol\Omega}$ must be \emph{antisymmetric}. 

More generally, starting from some arbitrary time $t$, we should have

\begin{align}
\tilde{\vec{U}}(t + \delta t) &\approx \left(\vec{I} + \tilde{\boldsymbol\Omega}\delta t\right)\tilde{\vec{U}}(t), \nonumber
\end{align}

\noindent Taking the limit $\delta t \to 0$, the time derivative of $\tilde{\vec{U}}(t)$ will then be

\begin{align}
\frac{d}{dt}\tilde{\vec{U}}(t) &= \tilde{\boldsymbol\Omega}\tilde{\vec{U}}(t), \nonumber
\end{align}

\noindent This may be solved by direct integration via the text book method of repeatedly back-substituting the resulting expressions for $\tilde{\vec{U}}(t)$ into the integrals on the right-hand-side. We then find the solution

\begin{align}
\tilde{\vec{U}}(t) &= \rme^{\tilde{\boldsymbol\Omega} t}. \label{eq:orthogonal-time-dependence}
\end{align}

\noindent (Note that we would obtain the same result if we had started from $d\tilde{\vec{U}}/dt = \tilde{\vec{U}}\tilde{\boldsymbol\Omega}$).

Now, since $\tilde{\boldsymbol\Omega}$ is antisymmetric, we will have 

\begin{align}
\det|\rme^{\tilde{\boldsymbol\Omega} t}| &= \rme^{\mathrm{Tr}[\tilde{\boldsymbol\Omega}]t} = \rme^{0} = 1, \nonumber
\end{align}

\noindent where $\mathrm{Tr}[\cdot]$ denotes the trace operator. Hence, any operator of the form of \eqref{eq:orthogonal-time-dependence} will be a member of the \emph{special orthogonal group} $SO(N)$.

\subsection{Consequences for $\mathbb{R}^{2}$ and $\mathbb{R}^{4}$}
Specifically considering the case of operators on a two dimensional real Hilbert space (as would be required to implement unary quantum gates) the group $SO(2)$ consists solely of matrices of the form

\begin{align}
\tilde{\vec{U}}(t) &= \left[\begin{array}{cc}
\cos\theta(t) & -\sin\theta(t) \\
\sin\theta(t) & \cos\theta(t) \end{array}\right],
\end{align}

\noindent for a counter-clockwise rotation of $\theta(t)$. None of the `real' gates $\vec{H}$, $\vec{X}$ or $\vec{Z}$ have this form, all of which have a determinant of -1, and so are \emph{not} part of $SO(2)$. Hence, \emph{no continuous time evolution can be given for these gates in terms of real operators on $\mathbb{R}^{2}$}. 

A similar problem pertains on $\mathbb{R}^{4}$ for the modelling of $4\times4$ gates acting on two qubits. The CNOT gate has a determinant of -1, so no member of $SO(4)$ can describe its evolution. We might also combine the steps involved in Bell state preparation to form the effective gate (actually the matrix of Bell state eigenvectors)

\begin{align}
\vec{U}_{Bell} = \frac{1}{\sqrt{2}}\left[\begin{array}{cccc}
1 & 0 & 1 & 0 \\
0 & 1 & 0 & 1 \\
0 & 1 & 0 & -1 \\
1 & 0 & -1 & 0 \end{array}\right], \label{eq:Bell}
\end{align}

\noindent This again has a determinant of -1, so no real Hilbert space formulation on $\mathbb{R}^{4}$ can model its continuous time dependence. 

Whether or not we can model the temporal evolution of quantum logic gates in a real Hilbert space of some higher dimension, what the results of this section show is that this cannot be done within a two dimensional real Hilbert space for rebits or a four dimensional real space for the gates acting on two rebits. However, as the reader may well be aware, this formulation is seldom attempted in earnest. Rather, the typical practice is to map the states of an $N$ dimensional complex space to a $2N$ dimensional real space. In the next section, we shall argue that this mapping actually replicates the structure of complex numbers and therefore just provides a `real' representation of complex states and operators.

\section{Mapping operators from $\mathbb{C}^{N}$ to $\mathbb{R}^{2N}$}\label{sec:mapping-CN-R2N}
\subsection{Isomorphism between scalar and matrix representations of complex numbers}
It is well-known that the scalar form of a complex number $z=x+y\rmi$ may be mapped isomorphically to a $2\times2$ matrix representation

\begin{align}
x + y\rmi &\leftrightarrow \left[\begin{array}{cc}
x & -y \\
y & x\end{array}\right]. \label{eq:complex-isomorphism}
\end{align}

\noindent In fact, there is also an alternative representation in which the signs on the off-diagonal matrix elements are reversed. It is important to realise that we have introduced nothing new here. Despite the fact that the matrix is specified in terms of the real numbers $x$ and $y$ (as is the scalar form $x + y\rmi$), this matrix \emph{is} a complex number - it is just a different way of representing a complex number without the explicit use of the symbol `$\rmi$'.

In fact, we may reintroduce the imaginary unit into the matrix representation by decomposing the RHS of \eqref{eq:complex-isomorphism} into its symmetric and antisymmetric components

\begin{align}
x + y\rmi &\leftrightarrow x\left[\begin{array}{cc}
1 & 0 \\
0 & 1\end{array}\right] + y\left[\begin{array}{cc}
0 & -1 \\
1 & 0\end{array}\right]. \nonumber 
\end{align}

\noindent We can then simplify this notation by defining

\begin{align}
\vec{I}_{2} &\equiv \left[\begin{array}{cc}
1 & 0 \\
0 & 1\end{array}\right]~\mathrm{and}~\vec{J}_{2} \equiv \left[\begin{array}{cc}
0 & -1 \\
1 & 0\end{array}\right], \nonumber 
\end{align}

\noindent where $\vec{I}_{2}$ plays the same role as the real number `1' and $\vec{J}_{2}$ substitutes for the imaginary unit `$\rmi$', having exactly the same additive and multiplicative properties. We now write

\begin{align}
x\times1 + y\times\rmi &\leftrightarrow x\times\vec{I}_{2}+ y\times\vec{J}_{2}, \nonumber 
\end{align}

\noindent which visually emphasises the equivalence.

It is pertinent to note that the matrix $\vec{J}_{2}$ will commute with a $2\times2$ matrix if and only if that matrix is of the form of the RHS of \eqref{eq:complex-isomorphism}. That is, \emph{if and only if that matrix represents a complex number}.

\subsection{Mapping general matrices}
Given the equivalence of \eqref{eq:complex-isomorphism}, one way of representing a complex $N\times N$ matrix is to replace every element with its $2\times2$ matrix equivalent. This then would produce a $2N\times 2N$ real matrix. However, it is `real' only in the sense that all its elements are specified by 2 real numbers. Due to their construction from $2\times2$ blocks, such matrices will still be imbued with a \emph{complex structure} not possessed by $2N\times 2N$ real matrices in general.

This method of replacing each complex element by its $2\times2$ matrix equivalent may appear rather cumbersome but may be achieved quite efficiently by decomposing a complex matrix into its real and imaginary parts and then taking the outer product with, respectively, the $\vec{I}_{2}$ and $\vec{J}_{2}$ matrices:

\begin{align}
\vec{A} &\to \mathrm{Re}[\vec{A}]\otimes\vec{I}_{2} + \mathrm{Im}[\vec{A}]\otimes\vec{J}_{2} \equiv \tilde{\vec{A}}. \label{eq:CN-to-R2N} 
\end{align}

\noindent This is, in fact, exactly the construction that most authors use when ostensibly constructing a real Hilbert space formulation of quantum mechanics. By construction, however, it is still a representation of a complex matrix. The test of this for an arbitrary $2N\times2N$ real matrix (perhaps constructed in some other way) is to check whether it commutes with the matrix

\begin{align}
\vec{J}_{2N} &\equiv \vec{I}_{N}\otimes\vec{J}_{2}, \label{eq:J2N} 
\end{align}

\noindent where $\vec{I}_{N}$ is the identity matrix on the original Hilbert space $\mathbb{C}^{N}$. This will give a block diagonal matrix with $\vec{J}_{2}$ as each $2\times2$ block. When multiplying an arbitrary $2N\times2N$ matrix $\tilde{\vec{A}}$ on a given side, each $2\times2$ block of $\tilde{\vec{A}}$ will be multiplied by on that side by  $\vec{J}_{2}$. Hence, $\tilde{\vec{A}}$ will commute with $\vec{J}_{2N}$ if and only if every $2\times2$ block of $\tilde{\vec{A}}$ commutes with $\vec{J}_{2}$. This would then mean that every such block represents a complex number. 

It should be noted that, here, we are using the `$\otimes$' operator to denote an \emph{outer} (or \emph{Kronecker}) product. More generally, the same symbol is often used to represent a \emph{tensor product}, for which the order of the operands holds no physical significance. For an outer product,  however, mathematically we have $\vec{A}\otimes\vec{B} \ne \vec{B}\otimes\vec{A}$. The two products do contain the same elements though - one is just a permutation of the other - so essentially contain the same information. 

It would have been equally viable, then, to perform the complex to real mapping via

\begin{align}
\vec{A} &\to \vec{I}_{2}\otimes\mathrm{Re}[\vec{A}] + \vec{J}_{2}\otimes\mathrm{Im}[\vec{A}]. \label{eq:CN-to-R2N-alt} 
\end{align}

\noindent as some authors do~\cite{aleksandrova2013real}. In this case, the test of whether a matrix was in fact a realisation of complex matrix would be to check its commutativity with

\begin{align}
\vec{J}_{2N}' &\equiv \vec{J}_{2}\otimes\vec{I}_{N}. \label{eq:J2N-prime} 
\end{align}

\noindent Clearly the two procedures are physically equivalent and the products of both can be shown to represent complex matrices. However, a test of a matrix produced using \eqref{eq:CN-to-R2N} may well fail if using the $\vec{J}_{2N}'$. Before any such test can be carried out, the matrix should be subject to the appropriate permutation transformation. If more than one matrices are being considered, the same permutation needs to be applied to all matrices for consistency. This will be an important consideration when we analyse the operators on a real vector space in Section~\ref{sec:antisym-R4}.

Interestingly, commutation with a matrix $\vec{J}$, having the properties of $\vec{J}_{2N}$ or a permutation of it,  was given by Stueckelberg~\cite{stueckelberg1960quantum} as a necessary condition for matrices representing observables in order to preserve the uncertainty principle in a real Hilbert space formulation. Whilst Stueckelberg's paper is seen as seminal in the area of real formulations of quantum mechanics, it is a little ironic that it arose out of lectures intended to explain to students why the imaginary unit was \emph{required} in quantum mechanics!

\subsection{Mapping unitary operators}\label{sec:mapping-unitary}
Let us consider the mapping of a unitary operator in an $N$-dimensional complex Hilbert space $\mathbb{C}^{N}$ to a matrix in $2N$-dimensional real Hilbert space $\mathbb{R}^{2N}$ via the prescription of \eqref{eq:CN-to-R2N}. Any such unitary operator in $\mathbb{C}^{N}$ can be written out according to its spectral decomposition as

\begin{align}
\vec{U} &= \boldsymbol\Phi\boldsymbol\Lambda\boldsymbol\Phi^{\dagger}, \nonumber
\end{align}

\noindent where $\boldsymbol\Lambda$ is an $N\times N$ diagonal matrix of the complex eigenvalues of $\vec{U}$ (with modulus of unity) and the columns of $\boldsymbol\Phi$ are the corresponding normalised eigenvectors. Mapping this to the real Hilbert space, we have

\begin{align}
\vec{U} &\to \tilde{\vec{U}} = \tilde{\boldsymbol\Phi}\tilde{\boldsymbol\Lambda}\tilde{\boldsymbol\Phi}^{T}, \nonumber
\end{align}

\noindent where now $\tilde{\boldsymbol\Lambda}$ is a $2N\times 2N$ block diagonal matrix containing the $2\times 2$ rotation matrices representing the complex eigenvectors of $\boldsymbol\Lambda$. Now, since we will have $\tilde{\boldsymbol\Phi}\tilde{\boldsymbol\Phi}^{T} = \vec{I}_{N}\otimes\vec{I}_{2} = \vec{I}_{2N}$, this will be a member of $SO(2N)$. Hence, the mapping of \eqref{eq:CN-to-R2N} sends every unitary matrix of $\mathbb{C}^{N}$ to a special orthogonal operator in $\mathbb{R}^{2N}$. These are precisely the operators required to model continuous time-evolution on a real Hilbert space. However, as we have seen, they are actually just constructed from the $2\times2$ matrix representation of complex numbers.

As an example of the mapping scheme described above, let us consider what the time evolution matrix of the $\vec{Z}$ gate looks like in $\mathbb{R}^{4}$.

\begin{align}
\tilde{\vec{Z}}(t) = \left[\begin{array}{cccc}
1 & 0 & 0 & 0 \\
0 & 1 & 0 & 0 \\
0 & 0 & \cos(\pi t/\tau) & -\sin(\pi t/\tau) \\
0 & 0 & \sin(\pi t/\tau) & \cos(\pi t/\tau) \end{array}\right]. \label{eq:real-Z}
\end{align}

\noindent It is easily verified that this is an orthogonal matrix with a determinant of 1 and so belongs to $SO(4)$. Moreover, we can see that each $2\times 2$ block that the matrix is built from represents the corresponding complex element of \eqref{eq:Z-time}.

\subsection{Mapping Hermitian operators}\label{sec:mapping-hermitian}
Whilst this paper is primarily concerned with unitary/orthogonal operators, we should not leave this section without discussing the all-important concept of Hermitian operators, which we use to model physical observables. A Hermitian operator has the property $\vec{H}^{\dagger} = \vec{H}$. For the mapping of \eqref{eq:CN-to-R2N}, conjugate transposition corresponds to simple transposition and we have

\begin{align}
\tilde{\vec{H}}^{T} &= \mathrm{Re}[\vec{H}]^{T}\otimes\vec{I}_{2}^{T} + \mathrm{Im}[\vec{H}]^{T}\otimes\vec{J}_{2}^{T}, \nonumber \\
&= \mathrm{Re}[\vec{H}]^{T}\otimes\vec{I}_{2} - \mathrm{Im}[\vec{H}]^{T}\otimes\vec{J}_{2}. \nonumber
\end{align}

\noindent Hence, in order to have $\tilde{\vec{H}}^{T} = \tilde{\vec{H}}$, we must have

\begin{align}
\mathrm{Re}[\vec{H}]^{T} &= \mathrm{Re}[\vec{H}]~\mathrm{and}~\mathrm{Im}[\vec{H}]^{T} = -\mathrm{Im}[\vec{H}]. \nonumber
\end{align}

\noindent That is, the real part of $\vec{H}$ must be \emph{symmetric} and the imaginary part \emph{antisymmetric}.

\section{The endomorphism of $\mathbb{R}^{2^{n}}$}\label{sec:antisym-R4}
In Section~\ref{sec:evolve-RN} it was shown that a real $N\times N$ matrix representing a continuous evolution operator must belong to the special orthogonal group $SO(N)$. It was then further shown that such operators take the form of \eqref{eq:orthogonal-time-dependence} - i.e. the exponential of an \emph{antisymmetric} matrix $\tilde{\boldsymbol\Omega}$. It was then demonstrated in Section~\ref{sec:mapping-unitary} that the unitary operators of a complex Hilbert space $\mathbb{C}^{N}$ are always mapped to special orthogonal operators acting on a real Hilbert space $\mathbb{H}^{2N}$. However, we showed that the mapping involved just constitutes the representation of a complex vector space by a real vector space with twice the dimensions.

An outstanding question then is can we go the other way? Do the special orthogonal operators of a general real Hilbert space actually \emph{always} represent complex matrices? Here, we investigate the vector space of the operators acting on a real Hilbert space with dimensions $N=2^{n}$, where $n$ is a positive integer. That is, the \emph{endomorphism} of $\mathbb{R}^{2^{n}}$,  $\mathrm{End}(\mathbb{R}^{2^{n}})$, the mapping of the vector space to itself. This is the appropriate real Hilbert space for modelling a system of $n-1$ qubits and their transformations. Aleksandrova \emph{et al}~\cite{aleksandrova2013real} argue that the place of the missing two degrees of freedom is taken up by a \emph{universal quantum bit}, which they interpret physically. We, of course, just argue that such real Hilbert spaces are in fact just representations of complex vector spaces and that the `universal bit' is just due to complex numbers being two dimensional.

Specifically, we shall show that the antisymmetric basis operators do all individually represent complex numbers (recall that antisymmetric operators are required to construct members of $SO(2^{n})$). However, since a given operator may be supported by several such basis elements, we can not always apply the same $\vec{J}_{2N}$ vector to all of them.

It does turn out, though, that the mappings such as \eqref{eq:CN-to-R2N} or \eqref{eq:CN-to-R2N-alt} (for systems of several qubits there will be other equally valid schemes) the resulting representation is \emph{always} a complex representation. This is due to the fact that these mappings do not make use of \emph{all} the possible operators of $\mathrm{End}(\mathbb{R}^{2^{n}})$. We argue that we should interpret this as meaning that workable real Hilbert space formulations are restricted to realisations of complex Hilbert spaces.

\subsection{Basis elements of $\mathrm{End}(\mathbb{R}^{2^{n}})$}
The space of the linear operators of vector space may be viewed as an abstract vector space in its own right, with a complete set of orthogonal basis elements $\vec{v}_{i}$ (here we drop the tilde notation as, even though we will usually chose them to be real valued, such elements are equally valid for both real and complex spaces). Any given matrix of a (real) vector space can then be written as the expansion

\begin{align}
\tilde{\vec{A}} &= \sum_{i} \vec{v}_{i}\langle\vec{v}_{i},\tilde{\vec{A}}\rangle_{F}, \label{eq:matrix-representation}
\end{align} 

\noindent where

\begin{align}
\langle\tilde{\vec{A}},\tilde{\vec{B}}\rangle_{F} &\equiv \mathrm{Tr}[\tilde{\vec{A}}^{\dagger}\tilde{\vec{B}}] \label{eq:frobenius-def}
\end{align} 

\noindent is the \emph{Frobenius inner product} of two matrices. This is directly analogous to the inner product of two vectors and plays the same role in giving us a mechanism to define the orthogonality of basis elements via

\begin{align}
\langle\vec{v}_{i},\vec{v}_{i'}\rangle_{F} =\delta_{ii'}. \label{eq:orthogonality-cond}
\end{align} 

\noindent Note that the `$\dagger$' sign implying conjugate transposition in the definition of \eqref{eq:frobenius-def} has just been used for generality and that for real matrices we can just replace this with simple transposition.

In the case of $n = 1$, so that the dimensions of the vector space are $N=2$, we can define $\vec{v}_{i} = \vec{e}_{i}/\sqrt{2}$ where

\begin{align}
\vec{e}_{0} \equiv \vec{I},~\vec{e}_{1} \equiv  \vec{X},~\vec{e}_{2} \equiv \vec{J}~\mathrm{and}~\vec{e}_{3} \equiv \vec{Z}. \label{eq:ei-basis}
\end{align}

\noindent Here, $\vec{I}$ and $\vec{J}$ are just the $2\times2$ identity and $\vec{J}_{2}$ matrices (dropping the 2 subscript for brevity), whilst $\vec{X}$ and $\vec{Z}$ are the Pauli $\sigma_{x}$ and $\sigma_{z}$ matrices.

For ease of reference, we note the multiplication table for each of these matrices is

\begin{align}
\begin{array}{c|cccc}
 & \vec{I} & \vec{X} & \vec{J} & \vec{Z} \\
 \cline{1-5}
 \rule{0pt}{3ex}
 \vec{I}^{T} & \vec{I} & \vec{X} & \vec{J} & \vec{Z} \\
 \vec{X}^{T} & \vec{X} & \vec{I} & \vec{Z} & \vec{J} \\
 \vec{J}^{T} & -\vec{J} & \vec{Z} & \vec{I} & \vec{X} \\
 \vec{Z}^{T} & \vec{Z} & -\vec{J} & -\vec{X} & \vec{I}
\end{array} \label{eq:basis-multi-table}
\end{align}

\noindent Noting further that $\mathrm{Tr}[\vec{X}] = \mathrm{Tr}[\vec{J}] = \mathrm{Tr}[\vec{Z}] = 0$ and $\mathrm{Tr}[\vec{I}] = 2$, it is straight-forward to see that, using the Frobenius inner product, these elements are all orthogonal to one another. The factor of $1/\sqrt{2}$ further ensures their normalisation so that the inner product of an element with itself is unity. 

Moving to the more general case for which the vector space has dimensions $N = 2^{n}$, we may define basis elements

\begin{align}
\vec{v}_{i} &\equiv \frac{1}{2^{n/2}}\vec{e}_{i_{n}}\otimes\ldots\otimes\vec{e}_{i_{1}}, \label{eq:R2n-basis}
\end{align}

\noindent where $i_{k} \in \{0, 1, 2, 3\}$ and the index $i$ may be conveniently given by

\begin{align}
i &= i_{n}\times4^{n-1} + \ldots + i_{1}\times4^{0}. \label{eq:i-index}
\end{align}

\noindent Note that in \eqref{eq:R2n-basis}, the `$\otimes$' operator specifically denotes the \emph{Kronecker product}.

To prove the orthogonality of the elements defined by \eqref{eq:R2n-basis}, we first note that, by the mixed product properties of the Kronecker product,

\begin{align}
\vec{v}_{i}^{T}\vec{v}_{i'} &= \frac{1}{2^{n}}(\vec{e}_{i_{n}}^{T}\otimes\ldots\otimes\vec{e}_{i_{1}}^{T})(\vec{e}_{i_{n}'}\otimes\ldots\otimes\vec{e}_{i_{1}'}), \nonumber \\
&=  \frac{1}{2^{n}}(\vec{e}_{i_{n}}^{T}\vec{e}_{i_{n}'})\otimes\ldots\otimes(\vec{e}_{i_{1}}^{T}\vec{e}_{i_{1}'}). \nonumber
\end{align}

\noindent Then, using the trace property of the Kronecker product, we have

\begin{align}
\mathrm{Tr}[\vec{v}_{i}^{T}\vec{v}_{i'}] &=  \frac{1}{2^{n}}\mathrm{Tr}[\vec{e}_{i_{n}}^{T}\vec{e}_{i_{n}'}]\times\ldots\times\mathrm{Tr}[\vec{e}_{i_{1}}^{T}\vec{e}_{i_{1}'}], \nonumber \\
&= \delta_{i_{n}i_{n}'}\times\ldots\times\delta_{i_{1}i_{1}'} = \delta_{ii'}, \nonumber
\end{align}

\noindent showing that these elements are both normalised and orthogonal to one another. Since each $\vec{e}_{i_{k}}$ element can take one of the four matrices of \eqref{eq:ei-basis}, this will constitute a set of $4^{n} = (2^{n})^{2}$ dimensions, which is the correct number of dimensions for the endomorphism of $\mathbf{R}^{2^{n}}$. Hence, this set is complete.

\subsection{Antisymmetric basis elements}
It should be clear to see that any antisymmetric basis element must have an \emph{odd} number of $\vec{J}$ matrices involved in its construction. Suppose such a $\vec{J}$ matrix occurs in the $k$th position of an antisymmetric vector

\begin{align}
\vec{v}_{i} &= \frac{1}{2^{n/2}}\vec{e}_{i_{n}}\otimes\ldots\otimes\vec{J}\otimes\ldots\otimes\vec{e}_{i_{1}},~(\vec{v}_{i} \in \mathcal{A}). \nonumber
\end{align}

\noindent We can then choose a permutation of the $\vec{J}_{2^{n}}$ matrix that places its $\vec{J}$ vector at the same position.  

\begin{align}
\vec{J}_{2^{n}}' &\equiv \vec{I}\otimes\ldots\otimes\vec{J}\otimes\ldots\otimes\vec{I}. 
\end{align}

\noindent This will clearly commute with the element $\vec{v}_{i}$ defined above

\begin{align}
\vec{J}_{2^{n}}'\vec{v}_{i} &= -\frac{1}{2^{n/2}}\vec{e}_{i_{n}}\otimes\ldots\otimes\vec{I}\otimes\ldots\otimes\vec{e}_{i_{1}} = \vec{v}_{i}\vec{J}_{2^{n}}'. \nonumber
\end{align}

\noindent This then tells us that every antisymmetric element $\vec{v}_{i}$ can be viewed as a real representation of a complex matrix.

However, in order to construct general antisymmetric matrices that commute with $\vec{J}_{2^{n}}'$, we would need to build it from antisymmetric elements that either had $\vec{J}$ or $\vec{I}$ in the same place, which will not be the case for \emph{all} such elements. 

Similarly, in order for $\vec{J}_{2^{n}}'$ to commute with a symmetric matrix, the elements building it would have to $\vec{I}$ or $\vec{J}$ in the same place (a symmetric element would, of course, have an even number of $\vec{J}$ vectors in its construction). Now, there will be symmetric operators that do not involve \emph{either} $\vec{I}$ or $\vec{J}$ that are composed entirely of $\vec{X}$ or $\vec{Z}$ vectors, neither of which commute with $\vec{J}$ (in fact they anti-commute). Therefore, there will be symmetric matrices which \emph{cannot} be viewed as being composed of complex numbers.

\subsection{The $\mathbb{C}^{2^{n-1}} \to \mathbb{R}^{2^{n}}$ mapping}
When mapping a complex vector space of dimension $2^{n-1}$ to a real one of dimension $2^{n}$, the procedure of \eqref{eq:CN-to-R2N} (or an equivalent mapping) does not do so one-to-one. This is self-evident since the endomorphism of  $\mathbb{C}^{2^{n-1}}$ has dimensions of $4^{n-1}$ whilst $\mathbb{R}^{2^{n}}$ has four times as many. 

In fact, the mapping only fills half of the operators of $\mathbb{R}^{2^{n}}$. Each basis element of $\mathbb{C}^{2^{n-1}}$ is only combined with $\vec{I}$ and $\vec{J}$ - the combinations with $\vec{X}$ and $\vec{Z}$ are omitted. However, since the $\vec{I}$ and $\vec{J}$ are always introduced in the same place (according to whatever convention one chooses), there will always be a \emph{single} $\vec{J}_{2^{n}}'$ vector that commutes with \emph{all} matrices so produced. This just reiterates our earlier conclusion that all such mappings constitute a real representation of complex operators. We may also note that if we form the union of the antisymmetric basis states generated from all possible mappings $\mathbb{C}^{2^{n-1}} \to \mathbb{R}^{2^{n}}$, this will be equal to the entire set of antisymmetric operators on $\mathrm{End}(\mathbb{R}^{2^{n}})$.

Only if we were to use the other operators of $\mathbb{R}^{2^{n}}$ that we do not map to, would we (possibly) be using mathematics that was not a representation of complex linear algebra. So long as the mapping is kept consistent with the original complex vector space, however, we are still using complex numbers.

\section{Conclusions}\label{sec:conclusions}
It should come as no particular surprise that the time evolution of quantum gates under the standard theory should yield complex intermediate states. This is hard-coded into the Schr{\"o}dinger equation, in which the time derivative is an imaginary operator. What we have explicitly shown, though, is that no rebit confined to a line of longitude on the Bloch sphere can stay on that line under the action of a physical gate. Moreover, we have shown that such temporal evolution is impossible to model on a two-dimensional real Hilbert space. In addition, we have seen that unitary operators on a complex Hilbert space must be mapped to \emph{special orthogonal} operators on a real Hilbert space with twice the dimensions. 

More generally, the usual prescription for mapping elements of a complex vector space to a real one with twice the dimensions is just a different representation of the original complex entities. To call these `real' matrices is entirely misleading. A truly \emph{real} representation of quantum mechanics should just rely on the mathematics of real numbers. 

Complex numbers are two dimensional, which provide two indispensable degrees of freedom via their modulus and phase. The squared modulus gives us the probabilistic interpretation via the Born rule, without which we would not be able to use quantum mechanics to make real world predictions of any kind. The time evolution emerges through the phase, without which we would have no dynamics. One then needs to ask, how would one dimensional real numbers provide this?  

\bibliographystyle{unsrt}
\bibliography{real-quantum}

%
%
%
%
%
%
%

\end{document}